\documentclass[fleqn,usenatbib]{mnras}

\usepackage{newtxtext,newtxmath}

\usepackage[T1]{fontenc}
\usepackage{ae,aecompl}

\usepackage{graphicx}	
\usepackage{amsmath}	
\usepackage{amssymb}	
\usepackage[xindy]{glossaries}
\usepackage{tabularx}
\glsdisablehyper
\usepackage{subcaption}
\usepackage{dcolumn}
\newcolumntype{d}[1]{D{.}{.}{#1}}

\newcommand{\sups}[1]{\textsuperscript{#1}}
\newcommand{\subs}[1]{\textsubscript{#1}}

\makeglossaries

\newacronym{dm}{DM}{Dispersion Measure}
\newacronym{frb}{FRB}{fast radio burst}
\newacronym{fwhm}{FWHM}{Full-Width at Half-Maximum}
\newacronym{htru}{HTRU}{High Time Resolution Universe}
\newacronym{snr}{S/N}{Signal-to-Noise Ratio}
\newacronym{vlbi}{VLBI}{Very Long Baseline Interferometry}

\title[Commensal FRB Survey]{GBTrans: A commensal search for radio pulses with the Green Bank twenty metre telescope}

\author[G. Golpayegani et al.]{Golnoosh Golpayegani$^{1,2}$\thanks{E-mail: gogolpayegani@mix.wvu.edu},
Duncan R. Lorimer$^{1,2}$,
Steven W. Ellingson $^{3}$, \newauthor
Devansh Agarwal$^{1,2}$,
Olivia Young$^{1,2}$,
Frank Ghigo $^{4}$,
Richard Prestage$^{1,2}$,\newauthor
Kaustubh Rajwade$^{5}$,
Maura A. McLaughlin$^{1,2}$
and Michael Mingyar$^{1,2}$
\\
$^{1}$Department of Physics and Astronomy, West Virginia University, Morgantown, WV 26505, USA\\
$^{2}$Center for Gravitational Waves and Cosmology, West Virginia University, Chestnut Ridge Research Building, Morgantown, WV 26505, USA\\
$^{3}$Bradley Department of Electrical and Computer Engineering, Virginia Tech, Blacksburg, VA 24061, USA\\
$^{4}$Green Bank Observatory, 155 Observatory Rd, Green Bank, WV 24944, USA\\
$^{5}$Jodrell Bank Centre for Astrophysics, University of Manchester, Oxford Road, Manchester M13 9PL, United Kingdom\\
}
\date{Accepted 2019 August 22. Received 2019 August 19; in original form 2019 May 01}

\pubyear{2019}
\hypersetup{draft}
\begin{document}
\label{firstpage}
\pagerange{\pageref{firstpage}--\pageref{lastpage}}
\maketitle

\begin{abstract}
We describe GBTrans, a real-time search system designed to find fast radio bursts (FRBs) using the 20-m radio telescope at the Green Bank Observatory. The telescope has been part of the Skynet educational program since 2015. We give details of the observing system and report on the non-detection of  FRBs from a total observing time of 503 days. Single pulses from four known pulsars were detected as part of the commensal observing. The system is sensitive enough to detect approximately half of all currently known FRBs and we estimate that our survey probed redshifts out to about 0.3 corresponding to an effective survey volume of around 124,000~Mpc$^3$. Modeling the FRB rate as a function of fluence, ${\cal F}$, as a power law with ${\cal F}^{-\alpha}$, we constrain the index $\alpha < 2.5$ at the 90\% confidence level. We discuss the implications of this result in the context of constraints from other FRB surveys.
\end{abstract}


\section{Introduction}
\label{sec:intro}

Pulsar searches and their need for high time and frequency resolution have opened new windows on the transient Universe. The best example of this so far is the discovery of fast radio bursts \citep[FRBs;][]{2007Sci...318..777L,2013Sci...341...53T}. FRBs are very bright transient radio pulses that occur on short ($\sim $ms) timescales, but emit about as much energy as the Sun produces in about a month. However, for the brightest known FRBs this number is about a day. At the time of writing, ninety FRBs have been published\footnote{http://www.frbcat.org} \citep[for an up-to-date list, see][]{2016PASA...33...45P}. Although this sample is currently not large enough to unambiguously characterize their origin and emission mechanism, it is clear that they form a cosmological population \citep[see, e.g.,][]{2016MNRAS.458..708C, 2017ApJ...834L...7T}. 

Though most FRBs have been detected as one-off events, a few of them have shown repetitions \citep{2016Natur.531..202S,2019Natur.566..235C,2019arXiv190803507T}. FRB~121102 was localized to a star-forming region in a dwarf galaxy with redshift 0.19, using the Karl G.~Jansky Very Large Array (VLA) acting jointly with single-dish observations using the 305-m William E.~Gordon Telescope at the Arecibo Observatory \citep{2017Natur.541...58C}. 

Follow-up studies showed a large and variable rotation measure towards this source, suggesting that FRB~121102 is in an extreme and dynamic magneto-ionic environment. A neutron star origin is consistent with both such an environment and the short burst durations \citep{2018Natur.553..182M}. 

The second repeating FRB, 180814.J0422+73, was recently discovered by CHIME \citep[Canadian Hydrogen Intensity Mapping Experiment][]{2019Natur.566..235C}. The CHIME collaboration reported the detection of six repeat bursts from FRB 180814.J0422+73 among 13 detected FRBs during the telescope's pre-commissioning phase \citep{2019Natur.566..230C}. Most recently, CHIME discovered eight repeating FRBs, varying from two repeating bursts to ten \citep{2019arXiv190803507T}. The variety of these recently detected repeaters is suggestive of different environmental properties (e.g.,~larger burst widths) and emission mechanisms for the repeating FRBs. Further discoveries with CHIME and other instruments are greatly anticipated.

Besides FRB~121102, two other single FRBs were recently localized using interferometry: The Australian Square Kilometre Array Pathfinder (ASKAP) interferometer, was able to localize FRB 180924 to a position 4~kpc from the center of a luminous galaxy \citep{2019arXiv190611476B}. This sub-arcsecond localization allowed them to identify the massive host galaxy with the redshift of 0.3214. FRB~190523 was detected with the Deep Synoptic Array ten-antenna prototype (DSA-10), which consists of 4.5-m radio
dishes separated by 6.75 m to 1300 m, located at the Owens Valley Radio Observatory. This FRB was also localized to a few-arcsecond region containing a single massive galaxy at a redshift of 0.66 which is the likely host galaxy of the burst \citep{2019arXiv190701542R}.

Over the past decade, many different hypotheses for the origin of FRBs have been suggested, from which some could be tested based on data from observed FRBs \citep{2016MPLA...3130013K}. The  main proposed models include giant pulses from pulsars \citep{2016MNRAS.457..232C,2016MNRAS.462..941L}, magnetar giant flares \citep{2017ApJ...843...84N,2017ApJ...841...14M}, merging or colliding neutron stars \citep{2016ApJ...822L...7W}, neutron star collapse \citep{2014A&A...562A.137F}, interaction of a pulsar with its environment \citep{2017ApJ...836L..32Z}, primordial black holes falling into neutron stars \citep{2018ApJ...868...17A}, and coalescing white dwarf binaries \citep{2041-8205-776-2-L39}.

\begin{figure*}
    \includegraphics[width=1.0\linewidth]{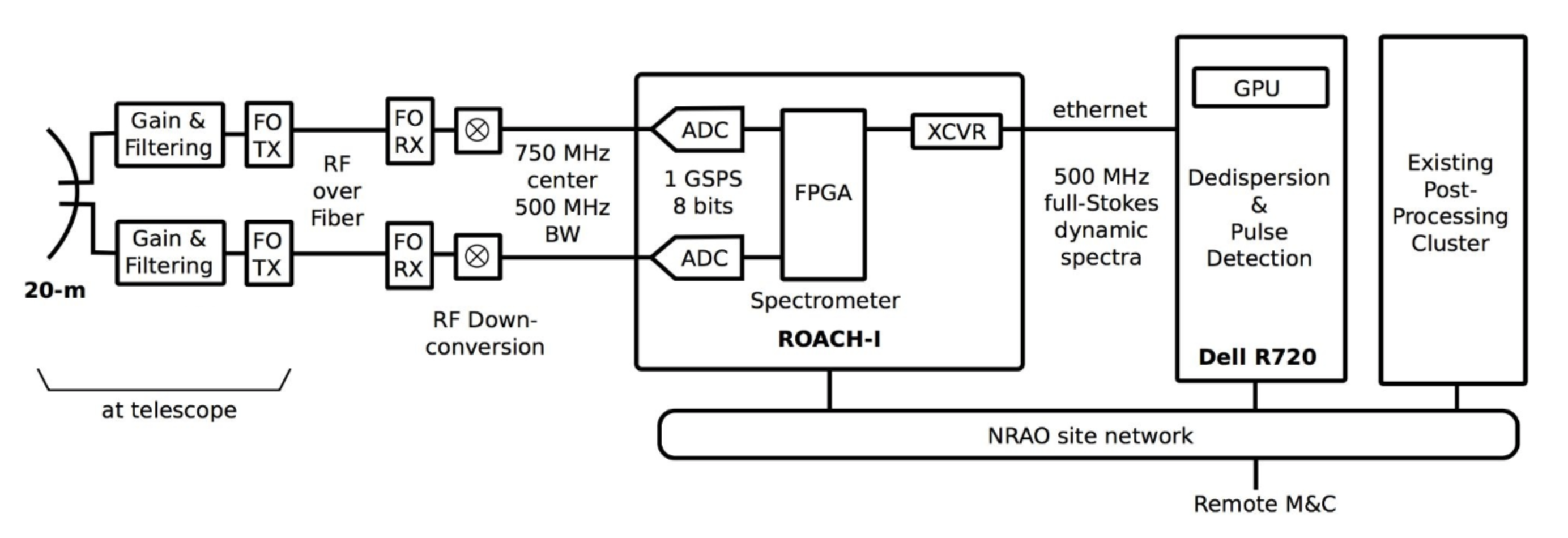}
    \caption{Block diagram showing the downstream electronics and data acquisition system summarising the existing system architecture developed for GBTrans.}
\end{figure*}

Many of the surveys carried out so far \citep[see their section 4.3 for a review]{2019A&ARv..27....4P} use telescopes with large collecting areas and high sensitivities, motivated by the relatively large flux densities of some FRBs
\citep[see, e.g.,][]{2018Natur.562..386S}, we have developed a real-time FRB detector on the 20-m telescope at the Green Bank Observatory, taking advantage of the extensive sky coverage available (approximately 80\% of the sky) and a large field of view of this smaller dish. This experiment, which we call GBTrans, is a synergistic effort partially supported from the Skynet Robotic Telescope Network Project\footnote{http://skynet.unc.edu}. The 20-m telescope
is also being used in a companion project (Gregg et al.~2019) which focuses on coordinated observations with the Neil Gehrels Swift Observatory.
The plan for the rest of this paper is as follows. In \S 2, we describe the GBTrans system and detection pipeline. In \S 3, we summarize the observations carried out and present the results of the survey including detected single pulses and giant pulses from known pulsars and candidate astrophysical pulses. In \S 4 we explain the method we used to estimate the FRB rate and survey volume for this survey and possible explanations for our non-detection of FRBs so far and speculate on future developments, and finally, in \S 5, we draw conclusions and make suggestions for future work.

\section{GBTrans description}

The 20-m telescope at the Green Bank Observatory in Green Bank, WV, has been in operation since late 1994. Originally funded by the US Naval Observatory, it was part of the National Earth Orientation Service telescope network, and participated in a global program of Earth Orientation VLBI (very long baseline interferometry) measurements in cooperation with the International Earth Rotation Service, and with the NASA Space Geodesy program. Following a shut down in 2000, the telescope was restored, automated, and made accessible as part of Skynet \citep{2016PASP..128e5002S,2013AAS...22134524H}. The main receiver currently in use operates at a central frequency of 1.4~GHz and provides a cryogenically cooled dual-polarization channel input for pulsar and spectral line work. Although there was some variation due to occasional cryogenics failures as well as telescope time spent on bright sources, the typical system temperature was found to be 40~K. The observing bandwidth is 80~MHz, centred at 1400~MHz.

A block diagram summarising the signal path from the sky to the data acquisition system developed for GBTrans is shown in Fig.~1. The signals are down-converted to a centre frequency of 750~MHz and digitized at 1~GHz before being converted to incoherent fully-polarimetric  dynamic spectra using a ROACH-I FPGA-based spectrometer. The spectrometer output is 2048 frequency channels with spectral resolution of
244~kHz and time resolution of $131 \mu$s, represented as 8-bit integers for all four Stokes parameters. The resulting data stream is slightly greater than 500~Mb/s, including meta-data. 

Real time analysis and detection is implemented on a GPU-equipped Dell R720 rack mount server
using purpose-built software developed by Virginia Tech. The server consists of dual Intel Xeon E5-2640 2.5~GHz 6-core CPUs, 32~GB RAM, $4 \times 1$ TB hot-pluggable hard drives, and an Nvidia Tesla K10 Graphics Processing Unit (GPU). Data analysis software is implemented in C and was developed to run on a Linux platform. The principal software components include a ring buffer, an executive processor, and a GPU-based processor. The ring buffer transfers data arriving synchronously from the spectrometer into shared memory, which allows the executive processor to operate asynchronously. The executive processor operates on arriving dynamic spectra in contiguous 13.1-s segments. Each segment is examined for data integrity (e.g., checking for correctly-ordered frame counters). As a diagnostic, spectra and total power for all four Stokes parameters are integrated over the segment and recorded. 

We take the 13.1-s data segments and use a GPU to produce de-dispersed total power time series using a brute-force algorithm for 531 trial dispersion measures (DMs) spanning the range 0--9900~cm$^{-3}$~pc. Each time series is subsequently box-car averaged in powers of 2 to search for single-sample pulses with widths in the range 131~$\mu$s to 268~ms. The resulting detection metrics are saved, and any  data segments containing pulses with signal-to-noise (S/N) ratios exceeding 10 and DM~$>10$ trigger a data-preservation protocol which causes a block of data to be written which we henceforth refer to as an event. Each event consists of the raw segment of full-Stokes data as well as all available meta-data which is saved on a post-processing cluster for long-term storage and follow-up analysis.

\section{Observations and data processing}
\label{sec:overview}

We have collected data with the aforementioned system from the beginning of December 2014 to the beginning of March 2018. Taking into account the days in which the system was down due to maintenance on the telescope or equipment failures, where no events were recorded, GBTrans was in operation for 503 days. Fig.~2 shows the distribution of events over the entire duration. 

\begin{figure}
    \includegraphics[width=1.0\linewidth]{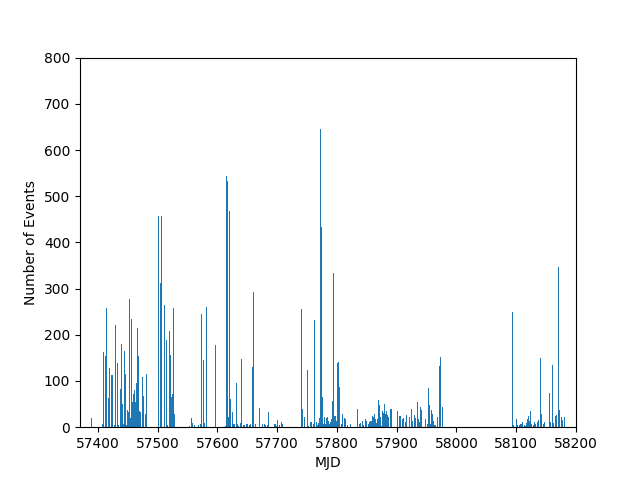}
    \caption{Histogram of the number of events versus date. There is very little data available before January 2016 (Seven epochs containing 27 events at MJDs from from 56998 to 57121). Most of our observations occurred between January 2016 and September 2017 (MJD range 57400---58000).}
\end{figure}

For each event, we applied a post-detection pipeline where the data were processed using the \textsc{heimdall}\footnote{https://sourceforge.net/projects/heimdall-astro} single-pulse software package. The generated candidates were clustered using the ``friends-of-friends'' algorithm \citep{1982ApJ...257..423H} in which groups of events were identified with the same DM within a tolerance of 20~cm$^{-3}$~pc, a time of arrival within 32 raw samples, and associated with an event of the highest S/N and pulse width. The resulting candidates were then appended to the output list and tested against the following criteria: Pulse widths shorter than 33.5~ms and S/N above 10. For each event that met these constraints, a diagnostic plot was generated which contained the original dynamic spectrum, the de-dispersed dynamic spectrum using the DM at which the pulse was detected with the highest S/N, along with a frequency collapsed time series of the detection with the length equal to twice the DM-dependant delay time, and were inspected visually. These were reviewed and categorised into three classes: known pulsar, radio-frequency interference and unidentified single pulses (i.e.~candidate FRBs). Five known pulsars were labelled as such by cross-correlating meta-data from the headers with
the \textsc{ATNF} pulsar catalogue
\citep{psrcat}\footnote{http://www.atnf.csiro.au/research/pulsar/psrcat}. The detection statistics for these pulsars
are summarized in Table 1. The other
two classes were reviewed and labelled manually. 
Although we did not detect any FRBs in this analysis, we did detect $\sim 22117$ giant pulses from the Crab pulsar which will be published elsewhere as part of a dedicated study of the Crab. An example Crab giant pulse is shown in Fig.~3. Since L-band is more problematic in terms of RFI, a significant amount of RFI events were present in our candidate output plots. In particular, 25115 false positive events, excluding the single pulses from known sources were detected. Fig.~4. shows an exotic RFI candidate example caused by a frequency-modulation continuous wave radar.

\begin{table}
    \begin{center}
    
    \begin{tabular}{ld{2.1}d{3.2}d{3.0}d{4.0}d{2.1}}
    
    \hline
    \multicolumn{1}{c}{PSR} & \multicolumn{1}{c}{S\subs{1400}} & \multicolumn{1}{c}{DM\subs{cat}}        & \multicolumn{1}{c}{N\subs{pulses}}   & \multicolumn{1}{c}{S/N\subs{max}} \\
                             & \multicolumn{1}{c}{(mJy)}        & \multicolumn{1}{c}{(cm\sups{--3}~pc)} &  &                       &     \\
    \hline
    
J0332+5434    &   203  &  26.76    &       344  &  13.05 \\
J0534+2200    &    14   &      56.77     &     22117      &     88.69   \\
J0835--4510    &  1050   &  67.97    &   13   & 10.51  \\
J1644--4459  & 296 & 478.80   &   318   &     34.38       \\ 
J2022+5154    &   27 &   22.55   &     1633  &  13.09 \\
    \hline
    
    \end{tabular}
    \end{center}
    
    \caption{Parameters for known pulsars detected by GBTrans. From left to right,
    we list pulsar name, mean flux density at 1400~MHz,
    catalogue DM, number of detected
    single-pulses, and maximum single-pulse S/N. The mean flux density at 1400~MHz and DM were obtained from the ATNF pulsar catalogue \citep{psrcat}.}
    \label{tab:knpsrtab}
\end{table}

\begin{figure}
    \includegraphics[width=1.0\linewidth]{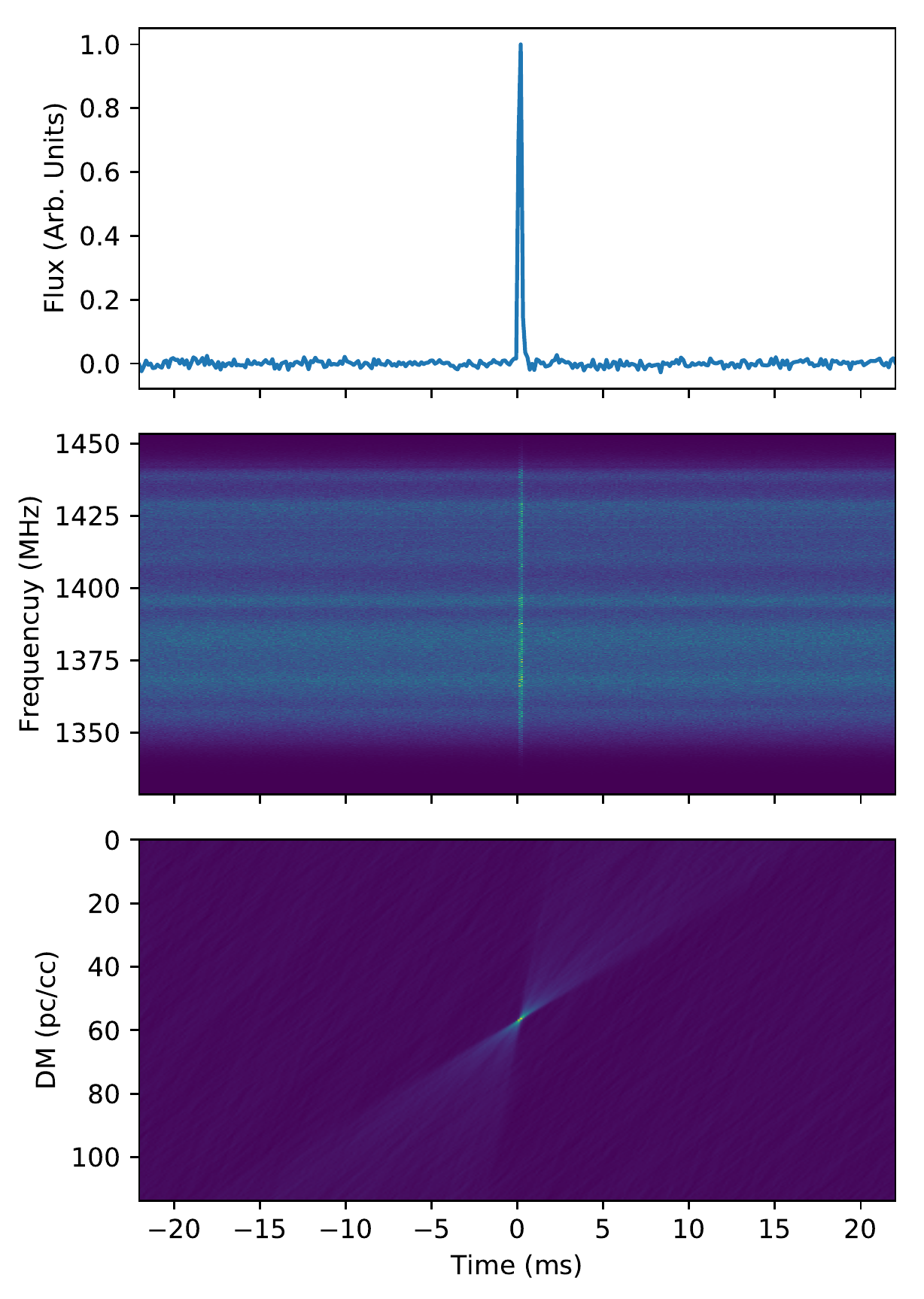}
    \caption{An example of a giant pulse (S/N~$\sim 90$) detected from the Crab pulsar, J0534+2200. The top panel shows the de-dispersed pulse in arbitrary flux units normalized to the peak intensity, the middle panel is the de-dispersed pulse in the frequency-time domain, and the bottom panel is the DM-time image. The 40~ms window is determined by doubling the DM-dependent dispersion delay of the pulse.}

\end{figure}

\begin{figure}
    \includegraphics[width=1.0\linewidth]{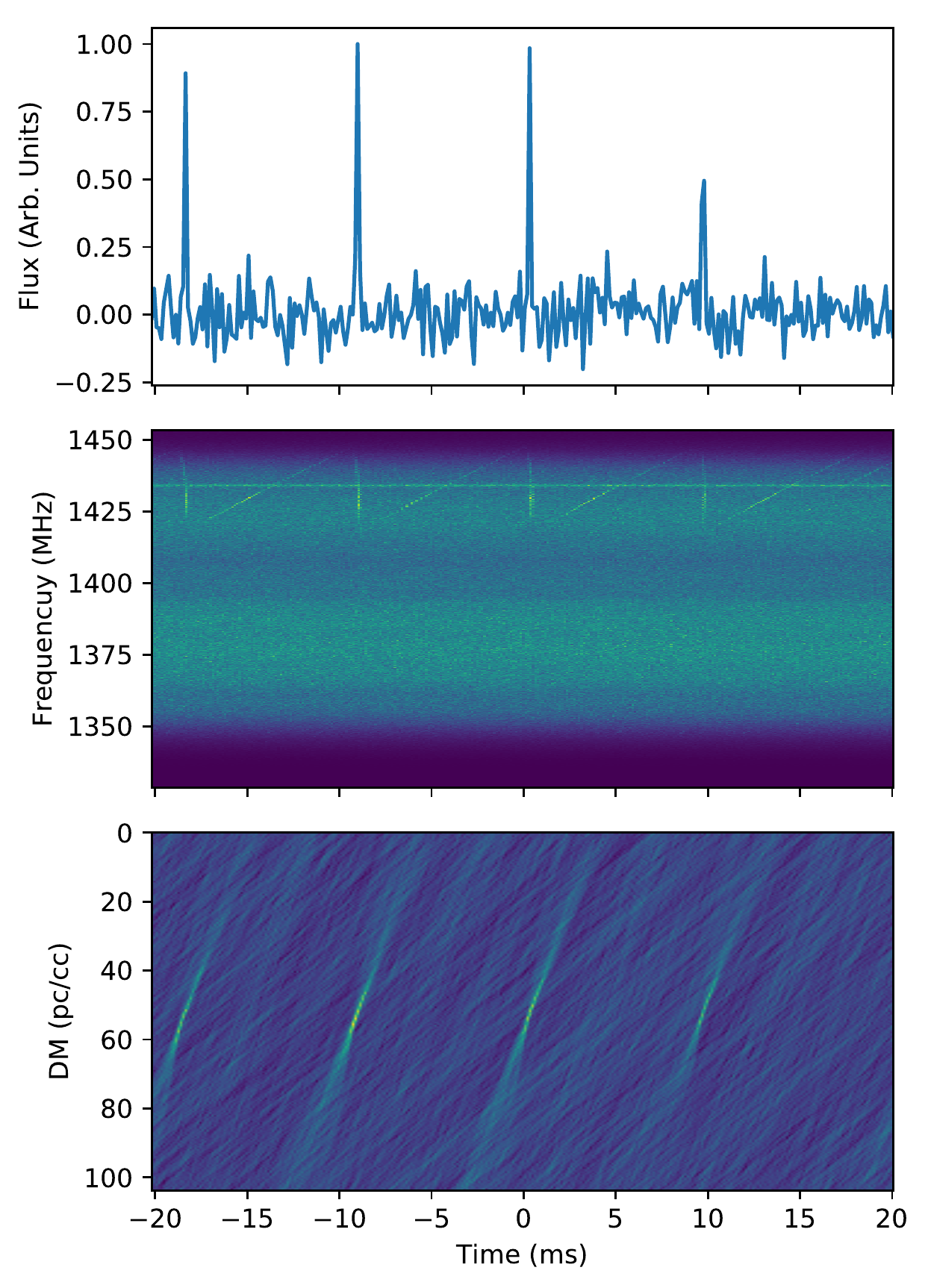}
    \caption{Example radio frequency interference signal caused by a frequency-modulation continuous wave radar. The top panel show the individual pulses, the middle panel is the de-dispersed pulses in frequency-time domain, and the bottom panel is the DM-time image.}
    
\end{figure}

\section{Discussion}
\label{sec:discuss}
\subsection{Expected FRB Rate}
When this experiment was being designed in 2013, the all-sky FRB rate, ${\cal R}$, was thought to be much higher than current estimates which are now based on larger samples of FRBs. Recent studies \citep[see, e.g.,][]{2017AJ....154..117L} now show that the event rate is almost an order of magnitude lower than
previously thought \citep[see, e.g.,][]{2013Sci...341...53T}. With this in mind, the lack of FRB detections in the survey, while disappointing, still provides useful constraints on the rate--fluence distribution.
In our analysis below, we first determine the instantaneous sensitivity and field of view of our experiment to FRBs. We then adopt a recent determination of the all-sky
FRB rate
${\cal R}_{\rm ASKAP} = 37 \pm 8$ bursts per sky per day with 1.4~GHz fluences 
above 26~Jy~ms which was found from an analysis of 
ASKAP detections
\citep{2018Natur.562..386S}
to determine realistic expectation times needed to make a detection.

To compute the sensitivity and sky coverage of GBTrans, we take the measured gain of the 20--m telescope, $G = A/2k$, where the effective surface area $A=237$~m$^{2}$ 
assumes an aperture efficiency of 75\% based on the modeled feed patterns, and $k$ is Boltzmann's constant. From this, we find $G = 0.086$~K~Jy$^{-1}$. Next, we use the radiometer equation \citep[see, e.g.,][]{2004hpa..book.....L} to compute for some limiting signal-to-noise ratio (S/N) the minimum detectable fluence
\begin{equation}
{\cal F}_{\rm min} = \frac{T_{\rm sys} \, {\rm S/N}}{G} \,
\sqrt{\frac{W}{2B}},
\end{equation}
where the typical system temperature $T_{\rm sys}=40$~K and the bandwidth $B=80$~MHz. For
consistency with the ASKAP survey, we adopt their minimum FRB pulse width $W=1.26$~ms, and 
a S/N threshold of 10.
This gives ${\cal F}_{\rm min} \simeq 6$~Jy~ms. The minimum detectable fluence at the full-width half maximum (FWHM) of the main beam of GBTrans is therefore about 12~Jy~ms. We adopt this as a fluence-complete limit in the analysis below. Fig.~5 shows the survey sensitivity among with previously detected FRBs. It appears that more than half of the current FRBs are detectable with GBTrans. Because our time resolution is better than for the ASKAP survey, as shown in Fig.~5, we also are sensitive to FRBs with widths shorter than 1~ms. Such FRBs do exist, and the observed sample is currently dominated by CHIME discoveries \citep{2019Natur.566..230C}. Because of the scaling between fluence and pulse width in Eq.~(1), the fluence limit at our sampling interval (131~${\mu}$s) is about 4~Jy~ms.

Having found the sensitivity out to the beam FWHM, we next need to compute the corresponding solid angle, $\Omega$, which represents the instantaneous amount of sky sampled at this limit. For a gaussian beam response
\cite[for a discussion, see][]{2016era..book.....C} we have $\Omega \simeq 1.133 \, {\rm FWHM}^2$, where for an observing
wavelength, $\lambda=0.2$~m,
\begin{equation}
{\rm FWHM} = 1.2 \frac{\lambda}{\sqrt{4A/\pi}} = 48'.
\end{equation}
From this, we find that the beam solid angle at the FWHM, $\Omega=2.2 \times 10^{-4}$~sr or $1.7 \times 10^{-5}$ of the whole sky. 

We model the rate--fluence distribution as a power law such that 
\begin{equation}
{\cal R(>{\cal F})} =
{\cal R}_{\rm ASKAP} \left(
\frac{ {\cal F} } {26 \, {\rm Jy\, ms}} \right)^{-\alpha},
\end{equation}
where the index $\alpha=1.5$
for Euclidean geometry. Keeping
$\alpha$ as a free parameter
but setting ${\cal F}={\cal F}_{\rm min}$, then we find an expression for the mean
``waiting time'', $T$, to detect a pulse. Since this is just the reciprocal of the rate scaled by the solid angle coverage, we find that
\begin{equation}
\label{eq:wait}
T = ({\cal R} \Omega)^{-1} = 
\left(\frac {1600 \pm 350 \, {\rm days}}
{2.2^{\alpha}}\right).
\end{equation}
In Fig.~\ref{fig:wait} we 
show Eq.~\ref{eq:wait} alongside
these various values of 
$\alpha$ from earlier
studies and our 
experimental limit on $T$. To be consistent with our experimental results, $T>503$ days. From this, as shown in Fig.~6, we estimate that $\alpha<1.7$.

Care should be taken when interpreting this simple point estimate of the upper limit because there is no confidence interval associated with it. To demonstrate this, assuming that FRBs as a population follow Poissonian statistics in their event rate, the probability of finding at least one
FRB in our data set $P_1=1-\exp(-{\cal R}\Omega T)$. Setting $\alpha=1.7$ in Eq.~3 to find ${\cal R}$ and $T=503$~days, we find $P_1=70\%$. To set a robust limit on $\alpha$, we can repeat this calculation to find $P_1$ as a function
of $\alpha$. Requiring $P_1 \ge 0.9$, we find that we
should have detected at least one FRB with 90\% confidence
if $\alpha>2.5$. We therefore conclude that
$\alpha<2.5$ at the 90\% confidence level.

Table~2 summarizes different $\alpha$ constraints reported in literature.
There is currently a wide range of $\alpha$ values that
are quoted.
An Euclidean rate--fluence distribution would therefore lead to $T \sim 1$~yr. \citet{2018MNRAS.474.1900M}
estimate, based on a
recent maximum likelihood
analysis on the Parkes FRBs,
that $\alpha = 2.6_{+1.3}^{-0.7}$.
For this range of $\alpha$ values,
we would expect waiting times in the range $58 < T < 436$~days.
In contrast, 
\citet{2017RAA....17....6L}
estimate $\alpha=0.14 \pm 0.20$.
This would correspond to
$956 < T < 2044$~days.

\begin{figure} \includegraphics[width=1.0\linewidth]{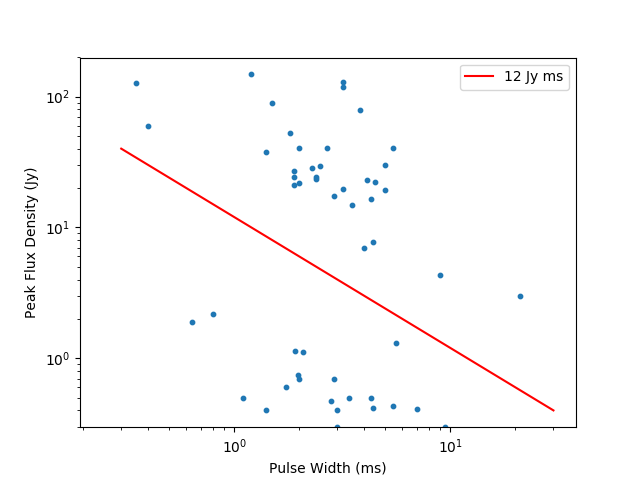}
\caption{Survey sensitivity of GBTrans based pulse width and peak flux density. The red line is the derived minimum detectable fluence. Blue dots are previously reported FRBs and events above this line are detectable with GBTrans.}
\label{fig:sw}
\end{figure}

\begin{figure} \includegraphics[width=1.0\linewidth]{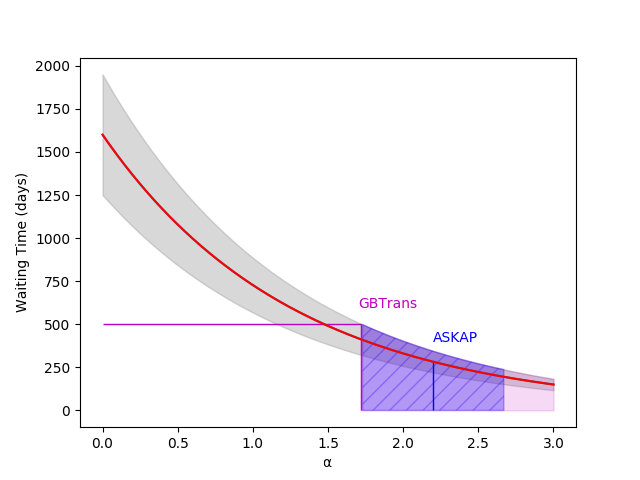}
\caption{Expected waiting time (red line) for GBTrans as a function of source-count index $\alpha$. 
The shaded grey region is the uncertainty of the function. The blue line is the $\alpha$ value estimation from ASKAP with its shaded blue uncertainty region \citep{2019MNRAS.483.1342J}. The pink line demonstrates the corresponding $\alpha$ value of this survey considering our non-detection after 503 days of observation. The shaded pink region is the area that is not consistent with our results.}
\label{fig:wait}
\end{figure}

\begin{table}
\begin{tabular}{l|l}
Source count index, $\alpha$ & Reference              \\ \hline
$2.6_{+1.3}^{-0.7}$   & \cite{2018MNRAS.474.1900M} \\
$2.2_{+0.6}^{-1.2}$ & \cite{2018MNRAS.475.1427B}    \\
$2.20 \pm 0.47$ (ASKAP) & \cite{2019MNRAS.483.1342J}  \\
$2.1_{+0.5}^{-0.6}$ & \cite{2018Natur.562..386S} \\
$<2.5$  & GBTrans --- this paper                \\
$1.18 \pm 0.24$ (Parkes) & \cite{2019MNRAS.483.1342J}       \\
0.91  & \cite{2017AJ....154..117L}    \\
0.8 -- 1.7 & \cite{2016MNRAS.461..984O}   \\
$0.9 \pm 0.3$  & \cite{2016MNRAS.458..708C} \\
0.5 -- 0.9  & \cite{2016ApJ...830...75V} \\
$0.14 \pm 0.20$ & \cite{2017RAA....17....6L}

\end{tabular}

\caption{Various estimates on $\alpha$ values ranked in descending order. Our constraint of $\alpha < 2$ is consistent with a number of estimates, as well as the expectation for a population of standard candles in Euclidean geometry in which $\alpha=1.5$.}
\end{table}

\subsection{Survey Volume}

Assuming a pulse width of 1~ms (consistent with the ASKAP FRB rate assumption), our nominal fluence limit discussed above corresponds to a peak flux limit of about 12~Jy. Adopting the standard candle model discussed by \citet{2013MNRAS.436L...5L} which gives a peak flux--redshift
relationship (see their Eq.~9), we find a maximum redshift reached by GBTrans to be approximately $z=0.3$. Given the beam solid angle computed in the previous section, the comoving volume corresponding to this limit assuming
a standard set of cosmological parameters for a flat universe
\citep{2014ApJ...794..135B} is 124,000~Mpc$^3$. As expected, this is substantially less than what \cite{2018MNRAS.474.3847F} reported ($z=3.3$ and 600,000~Mpc$^3$) for the more sensitive ALFABURST survey even with its smaller field of view. We note here that due to the limited sensitivity of GBTrans, the survey volume is significantly smaller than what we would naively expect from searching out to DM = 9900 pc cm$^{-3}$. For DMs dominated by intergalactic dispersion, the implied redshift for this DM is about $z=9$. Our choice of searching to such a high DM value is, therefore, very conservative but it mitigates against additional dispersion from FRBs buried in high-density environments with significant amounts of ionised plasma. Future discoveries of highly dispersed FRBs along with directly-measurable redshifts will be able to better quantify the local environments in these sources.

\section{Conclusions}
\label{sec:future_work}

GBTrans was an automated system that searched for FRBs commensally for over 500~days on a 20-m class telescope at 
Green Bank. The observations were  nominally sensitive to FRBs with redshifts out to about 0.3. Our non-detection during
this experiment leads to an upper limit on the power law index of the event rate--fluence 
exponent, $\alpha<2.5$ with 90\% confidence. With the torrent of discoveries expected from CHIME and ASKAP in the near future, the brightness distribution will undoubtedly be well probed by
these and other experiments. 

Our detection of numerous pulses from
known pulsars has validated the observing system. In addition to a
forthcoming publication concerning
giant pulses from the Crab pulsar found during the course of this project, future uses of the 20-m in the FRB field are migrating to targeted searches such as the {\it Swift} survey described in the companion paper 
by Gregg et al. Ongoing work aims to adapt the system to operate as a rapid response observer of radio transient signals associated with gamma-ray bursts.

\section*{ACKNOWLEDGEMENTS}

GBTrans activities are supported by a National Science Foundation (NSF) award AST-1516958. D.R.L. was supported by NSF award number OIA-1458952. G.G. would like to thank Mayuresh Surnis and Dan Werthimer for useful discussions and Nate Garver-Daniels for the technical support of this work. G.G., D.R.L., R.P. and M.A.M. are members of the NANOGrav Physics Frontiers Center which is supported by NSF award 1430284. S.W.E. was supported by NSF award AST-1516170 to Virginia Tech. K.R. would like to acknowledge financial support from European Research Council Horizon 2020 grant (no. 694745). The Green Bank Observatory is a facility of the National Science Foundation operated under cooperative agreement by Associated Universities, Inc. We thank the referee for very helpful comments on an earlier version of the manuscript.
\bibliographystyle{mnras}
\bibliography{gbtrans} 

\bsp	
\label{lastpage}
\end{document}